\documentclass[twoside,12pt]{article}
\usepackage{epsf}

\newcommand{\be}{\begin{eqnarray}}
\newcommand{\eq}[1]{eq.~(\ref{#1})}
\newcommand{\eqs}[2]{eqs.(\ref{#1},\ref{#2})}

\newcommand{\Eq}[1]{Eq.~(\ref{#1})}

\newcommand{\ur}[1]{(\ref{#1})}

\newcommand{\beq}{\begin{equation}}
\newcommand{\eeq}{\end{equation}}

\newcommand{\la}[1]{\label{#1}}
\newcommand{\bea}{\begin{eqnarray}}
\newcommand{\eea}{\end{eqnarray}}
\newcommand{\ba}{\begin{array}}
\newcommand{\ea}{\end{array}}

\newcommand{\half}{{\textstyle{\frac{1}{2}}}}
\newcommand{\at}{\overline{10}}
\newcommand{\oct}{\left({\bf 8},\frac{1}{2}\right)}
\newcommand{\dec}{\left({\bf 10},\frac{3}{2}\right)}
\newcommand{\adec}{\left({\bf\at},\frac{1}{2}\right)}
\newcommand{\pletthree}{\left({\bf 27},\frac{3}{2}\right)}
\newcommand{\pletone}{\left({\bf 27},\frac{1}{2}\right)}

\newcommand{\noi}{\noindent}
\newcommand{\n}{\nonumber}

\def\appendix{\par
\setcounter{subsection}{0}
\setcounter{equation}{0}
\def\thesection{Appendix}
\def\theequation{\Alph{section}.\arabic{equation}}}

\begin{document}
\thispagestyle{empty}
\vskip -1.5true cm
\hfill NORDITA-2003-59 NP

\vskip 1true cm

\begin{center}
{\Large\bf Exotic baryon multiplets at large number of colours}\\

\vskip .5true cm

{\large\bf Dmitri Diakonov$^{1,2}$ and Victor Petrov$^2$}\\

\vskip .5true cm

$^1$NORDITA, Copenhagen, Denmark\\
$^2$St. Petersburg Nuclear Physics Institute, Russia
\end{center}

\begin{abstract}
We generalize the usual octet, decuplet and exotic antidecuplet and
higher baryon multiplets to any number of colours $N_c$. We show that
the multiplets fall into a sequence of bands with $O(1/N_c)$ splittings
inside the band and $O(1)$ splittings between the bands characterized
by ``exoticness", that is the number of extra quark-antiquark pairs needed
to compose the multiplet. Each time one adds a pair the baryon mass is
increased by the same constant which can be interpreted as a
mass of a quark-antiquark pair. At the same time, we prove that masses of
exotic rotational multiplets are reliably determined at large $N_c$ from
collective quantization of chiral solitons.
\end{abstract}

\section{Introduction}
The successful theoretical prediction of a relatively light and narrow
exotic baryon $\Theta^+$ \cite{97} and its subsequent experimental
observation \cite{Osaka,ITEP,JLab,ELSA,Asratyan}  have stimulated much
interest in the dynamics of baryons that cannot be made of three
quarks.

$\Theta^+$ has been predicted from a chiral soliton view on
baryons, implying a semiclassical approximation. The approximation is
justified at large $N_c$, although one puts $N_c\!=\!3$ at the end of the
calculations.  At large $N_c$ baryons {\em are} chiral solitons of some
effective chiral action \cite{Witten}. The question is whether this approach
can give reliable description of the real $N_c\!=\!3$ world. By reliable we
mean that physical quantities like masses, widths, splittings etc., can
be computed in the large-$N_c$ limit, with known or at least controllable
$1/N_c$ corrections.  Much of this work has been done before for non-exotic
baryons; our aim is to extend it to exotic baryons that cannot be composed
of $N_c$ quarks.

In order to understand the scaling of baryon properties with $N_c$ one has
to construct explicitly  $SU(3)$ flavour multiplets (or representations)
that are arbitrary-$N_c$ prototypes of the lightest baryon
multiplets -- the octet with spin one half $\oct$, the decuplet with spin
three halves $\dec$, the antidecuplet with spin one half $\adec$, etc. We
generalize the previous study of this subject by Dulinski and Praszalowicz
\cite{DuPrasz} and Cohen\cite{Cohen}. We classify the multiplets at
arbitrary $N_c$ by ``exoticness" -- the minimal number of extra
quark-antiquark pairs one needs to add to the usual $N_c$ quarks to build
the multiplet. All multiplets that we shall be discussing appear as
rotational excitations of a chiral soliton. We compute their energies and
observe that the spectrum is equidistant in exoticness: each time one adds a
pair it costs, at large $N_c$, a fixed energy independent of $N_c$. Being
quite natural from the constituent quark point of view, this result is
somewhat unusual for a rotational spectrum. Moreover, very recently Cohen
\cite{Cohen} has expressed doubt whether the collective-quantization
description of a particular exotic multiplet $\adec$ (to which the newly
discovered $\Theta^+$ presumably belongs) makes sense. In the second
part of the paper we show that this doubt is ungrounded.
Furthermore, we prove that the masses of exotic multiplets, obtained from
the rotational spectrum, get only small $O(1/N_c)$ corrections from
mixing with other, non-rotational degrees of freedom.

\section{Generalization of  baryon multiplets to arbitrary $N_c$}

\begin{figure}
\begin{center}
\begin{picture}(100,120)
\put(-100,-180) {
\epsfxsize=9cm
\epsfysize=12cm
\epsfbox{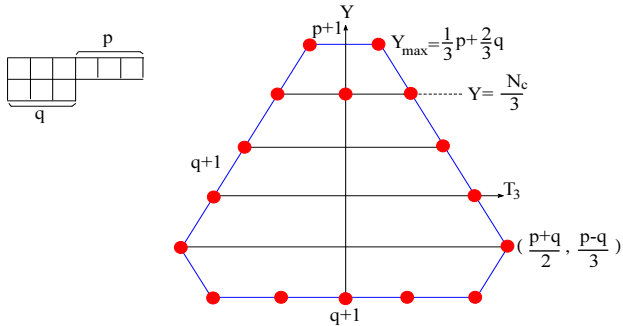} }
\end{picture}
\end{center}
\caption{A weight diagram of a generic $SU(3)$ representation in the
$(T_3,Y)$ plane.}
\label{genericSU3}
\end{figure}

A generic $SU(3)$ multiplet (or irreducible representation) is uniquely
determined by two non-negative integers $(p,q)$ having the meaning
of upper (lower) components of the irreducible $SU(3)$ tensor
$T^{\{f_1...f_p\}}_{\{g_1...g_q\}}$ symmetrized both in upper and lower
indices and with a contraction with any $\delta^{g_n}_{f_m}$ being zero.
Schematically, $q$ is the number of boxes in the lower line of the
Young tableau depicting an $SU(3)$  representation and $p$ is the
number of extra boxes in its upper line. The dimension of a
representation (or the number of particles in the multiplet) is

\beq
{\rm Dim}(p,q)=(p+1)(q+1)\left(1+\frac{p+q}{2}\right)
\la{Dim}\eeq
and the eigenvalue of the quadratic Casimir operator is given by

\beq
C_2(p,q)=\frac{1}{3}\left[p^2+q^2+pq+3(p+q)\right].
\la{C2}\eeq
The highest weight of a $(p,q)$ representation ({\it i.e.} the state
that generates all other states in the multiplet by applying
`eigenvalue-lowering' operators) is given by

\beq {\bf e}_{\rm H}(p,q)
=\left(T_3\!=\!\frac{p+q}{2},\,Y\!=\!\frac{p-q}{3}\right)
\la{mH}\eeq where $T_3$ is the third projection of the isospin and
$Y$ is the hypercharge. On the weight $(T_3,Y)$ diagram, a generic
$SU(3)$ representation is depicted by a hexagon, whose upper
(horizontal) side contains $p+1$ `dots' or particles, the adjacent
sides contain $q+1$ particles, with alternating $p+1$ and $q+1$
particles in the rest sides, the corners included --  see Fig.~1.
If either $p$ or $q$ is zero, the hexagon reduces to a triangle.
Particles on the upper (horizontal) side of the hexagon have
hypercharge

\beq
Y_{\rm max}=\frac{1}{3}p+\frac{2}{3}q
\la{Ymax}\eeq
being the maximal possible hypercharge of a multiplet with given
$(p,q)$.

The quantization of the chiral soliton rotation in flavour and ordinary
spaces proceeds as follows \cite{Guad,DP7,Prasz,Chemtob,ind}.
The lagrangian of the $SU(3)$ rotations is

\beq
{\cal L}_{\rm rot} =
\frac{I_1}{2}\left(\Omega_1^2+\Omega_2^2+\Omega_3^2\right)
+ \frac{I_2}{2}\left(\Omega_4^2+\Omega_5^2 +\Omega_6^2+\Omega_7^2\right)
- \frac{N_c}{2\sqrt{3}}\Omega_8
\la{Lrot8}\eeq
where $\Omega_A$ are angular velocities of the soliton
and $I_{1,2}$ are the two soliton moments of inertia, depending on its
concrete dynamical realization. Rotation along the 8th axis in flavour space
commutes with the `upper-left-corner' Ansatz for the soliton field,
therefore there is no quadratic term in $\Omega_8$. However there is a
Wess--Zumino--Witten term linear in $\Omega_8$. The canonical quantization
leads to the hamiltonian

\beq
{\cal H}_{\rm rot}=\frac{J_1^2+J_2^2+J_3^2}{2I_1}+
\frac{J_4^2+J_5^2+J_6^2+J_7^2}{2I_2},
\la{Hrot}\eeq
where the angular momenta satisfy the $SU(3)$ commutation relations.
\Eq{Hrot} must be supplemented by the quantization condition $J_8
= -N_c/2\sqrt{3} = -Y^\prime\sqrt{3}/2$ following from the
Wess--Zumino--Witten term. Given that

\beq
\sum_{A=1}^3J_A^2=J(J+1),\qquad
\sum_{A=1}^8J_A^2=C_2(p,q),\qquad J_8^2=\frac{N_c^2}{12},
\la{sums}\eeq
one gets the rotational energy of baryons  with given spin $J$
and belonging to representation $(p,q)$:

\beq
{\cal E}_{\rm rot}(p,q,J)=
\frac{C_2(p,q)-J(J+1)-\frac{N_c^2}{12}}{2I_2}+\frac{J(J+1)}{2I_1}.
\la{Erot}\eeq
Only those multiplets are realized as rotational excitations that have
members with hypercharge $Y=\frac{N_c}{3}$; if the number of particles with
this hypercharge is $n$ the spin $J$ of the multiplet is such that $2J+1=n$.
It is easily seen that the number of particles with a given $Y$
is $\frac{4}{3}p+\frac{2}{3}q+1-Y$ and hence the spin of the
allowed multiplet is

\beq
J= \frac{1}{6}(4p+2q-N_c).
\la{J}\eeq
A common mass ${\cal M}_0$ must be added to \eq{Erot} to
get the mass of a particular multiplet. Throughout this paper we are
disregarding the splittings inside multiplets as due to non-zero current
strange quark mass.

The condition that a horizontal line $Y=\frac{N_c}{3}$ must be inside
the weight diagram for the allowed multiplet leads to the requirement

\beq
\frac{N_c}{3}\leq Y_{\rm max} \qquad {\rm or} \quad p+2q\geq N_c
\la{cond}\eeq
showing that at large $N_c$ multiplets must have a high dimension.

We introduce a non-negative number which we name ``exoticness'' $E$ of a
multiplet defined as

\beq
Y_{\rm max}\;=\;\frac{1}{3}p+\frac{2}{3}q\;\equiv\; \frac{N_c}{3}+E, \qquad
E\geq 0.
\la{E}\eeq
Combining \eqs{J}{E} we express $(p,q)$ through $(J,E)$:
\bea
\n
p &=& 2 J - E, \\
\la{pq_JE} q &=& \frac{1}{2}N_c +2E-J. \eea
The total number of
boxes in Young tableau is $2q+p=N_c+3E$. Since we are dealing with
unit baryon number states, the number of quarks in the multiplets
we discuss is $N_c$, {\em plus} some number of quark-antiquark
pairs. In the Young tableau, quarks are presented by single boxes
and antiquarks by double boxes. It explains the name
``exoticness":  $E$ gives the minimal number of additional quark-antiquark
pairs one needs to add on top of the usual $N_c$ quarks to compose
a multiplet.

\begin{figure}[t]
\begin{center}
\begin{picture}(100,180)
\put(-240,-350) {
\epsfxsize=20cm
\epsfysize=20cm
\epsfbox{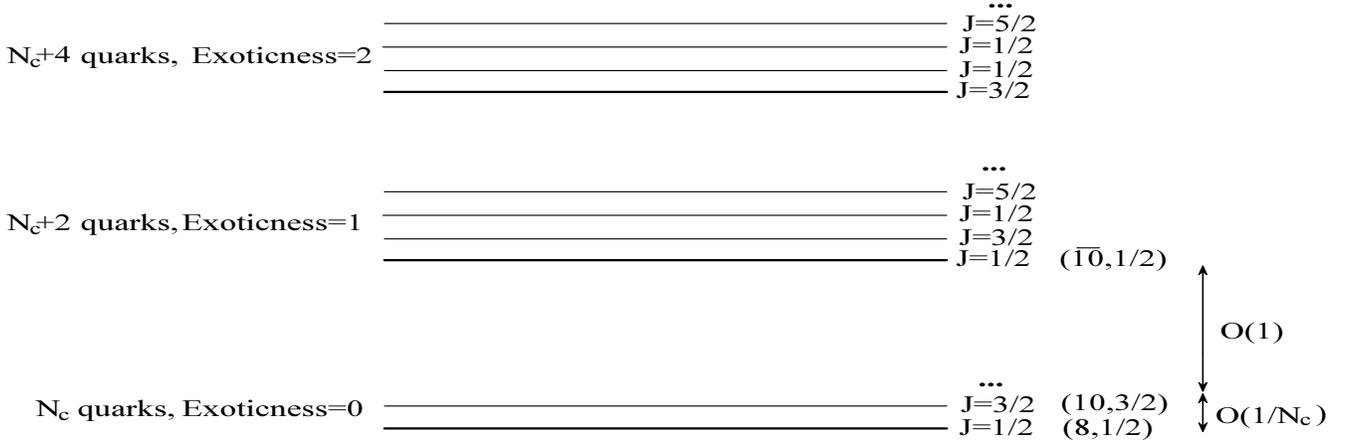} }
\end{picture}
\end{center}
\caption{Rotational excitations form a sequence of bands.}
\label{spectrumSU3}
\end{figure}

Putting $(p,q)$ from \eq{pq_JE} into \eq{Erot} we obtain the rotational
energy of a soliton as function of the spin and exoticness of the multiplet:

\beq
{\cal E}_{\rm rot}(J,E)=\frac{E^2+E(\frac{N_c}{2}+1-J)+\frac{N_c}{2}}{2I_2}
+\frac{J(J+1)}{2I_1}.
\la{ErotJE}\eeq
We see that for given $J\leq \frac{N_c}{2}+1$ the multiplet mass is a
monotonically growing function of $E$: the minimal-mass multiplet has
$E=0$. Masses of multiplets with increasing exoticness are:
\bea
\la{E0}
{\cal M}_{{\rm E}\!=\!0}(J) &=&
{\cal M}_0^\prime+\frac{J(J+1)}{2I_1},\qquad {\rm where}\quad
{\cal M}_0^\prime\equiv {\cal M}_0+\frac{N_c}{4I_2}, \\
\n\\
\la{E1}
{\cal M}_{{\rm E}\!=\!1}(J) &=&
{\cal M}_0^\prime+\frac{J(J+1)}{2I_1}+1\cdot \frac{\frac{N_c}{2}+2-J}{2I_2}
, \\
\n\\
\la{E2}
{\cal M}_{{\rm E}\!=\!2}(J) &=&
{\cal M}_0^\prime+\frac{J(J+1)}{2I_1}+2\cdot
\frac{\frac{N_c}{2}+2-J}{2I_2}+\frac{1}{I_2},\\
\n\\
\la{E3}
{\cal M}_{{\rm E}\!=\!3}(J) &=&
{\cal M}_0^\prime+\frac{J(J+1)}{2I_1}+3\cdot
\frac{\frac{N_c}{2}+2-J}{2I_2}+\frac{3}{I_2},
\qquad {\rm etc.}
\eea
At this point it should be recalled that both moments of inertia
$I_{1,2}=O(N_c)$, as is ${\cal M}_0$. We see from eqs.\ur{E0}-\ur{E3}
that multiplets fall into a sequence (labelled by exoticness) of rotational
bands with small $O(1/N_c)$ splittings inside the bands; the separation
between bands with different $E$ is $O(1)$. The corresponding masses
are schematically shown in Fig.~2.

The lowest band is non-exotic ($E\!=\!0$); the multiplets are
determined by $(p,q)=\left(2J,\frac{N_c}{2}-J\right)$, and their
dimension is $\;Dim=(2J+1)(N_c+2-2J)(N_c+4+2J)/8$ which in the
particular (but interesting) case of $N_c=3$ becomes 8 for spin
one half and 10 for spin 3/2. These are the correct lowest
multiplets in real world, and the above multiplets are their
generalization to arbitrary values of $N_c$. To make baryons
fermions one needs to consider only odd $N_c$.

Recalling that $u,d,s$ quarks' hypercharges are 1/3, 1/3 and -2/3,
respectively, one observes that all baryons of the non-exotic
$E\!=\!0$ band can be made of $N_c$ quarks. The upper side of
their weight diagrams (see Fig.~3) is composed of $u,d$ quarks
only; in the lower lines one consequently replaces $u,d$ quarks by
the $s$ one. This is how the real-world $\oct$ and $\dec$
multiplets are arranged and this property is preserved in their
higher-$N_c$ generalizations. The construction coincides with that
of ref. \cite{DuPrasz}. At high $N_c$ there are further multiplets
with spin  5/2 and so on. The maximal possible spin at given $N_c$
is $J_{\rm max}=\frac{N_c}{2}$: if one attempts higher spin, $q$
becomes negative.

\begin{figure}[t]
\begin{center}
\begin{picture}(100,70)
\put(-120,-15) {
\epsfxsize=12cm \epsfysize=3.5cm
\epsfbox{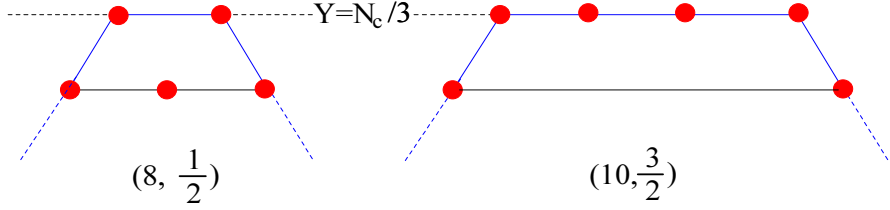} }
\end{picture}
\end{center}
\caption{Non-exotic ($E\!=\!0$) multiplets that can be composed of $N_c$
quarks.}
\label{exot_0}
\end{figure}

The rotational bands for  $E\!=\!1$ multiplets are shown in
Fig.~4.  The upper side of the weight diagram is exactly one unit higher
than the line $Y=\frac{N_c}{3}$ which is non-exotic, in the sense that its
quantum numbers can be, in principle,  achieved from exactly $N_c$ quarks.
However, particles corresponding to the upper side of the weight diagram
cannot be composed of $N_c$ quarks but require at least one additional
$\bar s$ quark and hence {\em one additional quark-antiquark pair} on top of
$N_c$ quarks.

\begin{figure}[b]
\begin{center}
\begin{picture}(100,100)
\put(-110,-15)
{
\epsfxsize=9cm
\epsfysize=3.5cm
\epsfbox{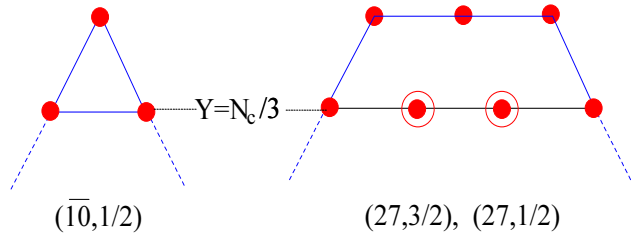} }
\end{picture}
\end{center}
\caption{Exotic ($E\!=\!1$) multiplets that can be composed of $N_c$ quarks
and one extra pair.}
\label{exot_1}
\end{figure}

The multiplet shown in Fig.~4, left, has only one particle with
$Y=\frac{N_c}{3}+1$. It is an isosinglet with spin $J\!=\!\frac{1}{2}$, and
in the quark language is built of $(N_c+1)/2$ $ud$ pairs and one $\bar s$
quark. It is the generalization of the $\Theta^+$ baryon to arbitrary odd
$N_c$. As seen from \eqs{Dim}{pq_JE}, the multiplet to which the
``$\Theta^+$" belongs is characterized by $(p,q)=\left(0,(N_c+3)/2\right)$,
its dimension is $(N_c+5)(N_c+7)/8$ becoming the $\adec$ at $N_c\!=\!3$.
Its splitting with the $N_c$ generalization of the non-exotic $\oct$
multiplet follows from \eq{E1}:

\beq
{\cal M}_{\at,\frac{1}{2}}-{\cal M}_{8,\frac{1}{2}}=\frac{N_c+3}{4I_2}
\la{antiten_oct}\eeq
coinciding with the recent finding of ref. \cite{Cohen}. Here and in what
follows we denote baryon multiplets by their dimension at $N_c\!=\!3$
although at $N_c\!>\!3$ their dimension is higher, as given by \eq{Dim}.

The second rotational state of the $E\!=\!1$ sequence has $J=\frac{3}{2}$;
it has $(p,q)=(2,(N_c+1)/2)$ and dimension $3(N_c+3)(N_c+9)/8$ reducing to
the mutliplet $\pletthree$ at $N_c=3$, see Fig.~4, right. In fact there
are two physically distinct multiplets there. Indeed, the weights in the
middle of the second line from top on the weight diagram  (with
$Y=\frac{N_c}{3}$) are twice degenerate, corresponding to spin 3/2 and 1/2.
Therefore, there is another  $3(N_c+3)(N_c+9)/8$-plet  with unit exoticness,
but with spin 1/2. At $N_c\!=\!3$ it reduces to $\pletone$. The splittings
with non-exotic multiplets are
\bea
\la{273}
{\cal M}_{27,\frac{3}{2}}-{\cal M}_{10,\frac{3}{2}}
&=&\frac{N_c+1}{4I_2},\\
\n\\
\la{271}
{\cal M}_{27,\frac{1}{2}}-{\cal M}_{8,\frac{1}{2}}
&=&\frac{N_c+7}{4I_2}.
\eea
The $E\!=\!1$ band continues to the maximal spin $J_{\rm max}=(N_c+4)/2$
where $q$ becomes zero.

These higher multiplets in the rotational spectrum of the $SU(3)$ soliton at
 $N_c=3$ has been known to the skyrmion community from the 1980's. After the
discovery of $\Theta^+$ there has been a renewed interest in them
\cite{KW,Kob}.

The $E\!=\!2$ rotational band starts from two states with spin 3/2 and 1/2
both belonging to the $SU(3)$ representation
$(p,q,Dim)=\left(1,(N_c+5)/2,(N_c+7)(N_c+11)/4\right)$. It reduces to the
${\overline{\bf 35}}$ multiplet at $N_c\!=\!3$. Their splittings with
non-exotic multiplets are
\bea
\la{353}
{\cal M}_{\overline{35},\frac{3}{2}}-{\cal M}_{10,\frac{3}{2}}
&=&\frac{N_c+3}{2I_2},\\
\n\\
\la{351}
{\cal M}_{\overline{35},\frac{1}{2}}-{\cal M}_{8,\frac{1}{2}}
&=&\frac{N_c+6}{2I_2}.
\eea
The maximal spin of the $E\!=\!2$ rotational band is
$J_{\rm max}=(N_c+8)/2$.

\begin{figure}
\begin{center}
\begin{picture}(100,100)
\put(-70,-15)
{
\epsfxsize=7cm
\epsfysize=4.5cm
\epsfbox{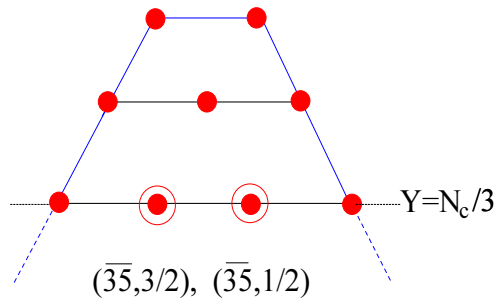} }
\end{picture}
\end{center}
\caption{Two lowest exotic ($E\!=\!2$) multiplets that can be composed of
$N_c$ quarks and two extra pairs.}
\label{exot_2}
\end{figure}

The upper side in the weight diagram (Fig.~5) for the $E\!=\!2$ sequence
has hypercharge $Y_{\rm max}=\frac{N_c}{3}+2$. Therefore, one needs
{\em two} $\bar s$ quarks to get that hypercharge and hence the multiplets
can be minimally constructed of $N_c$ quarks plus {\em two additional
quark-antiquark pairs}. This explains the name ``exoticness" $E$: it gives
the minimal number of additional quark-antiquark pairs needed to construct a
multiplet, on top of the usual $N_c$ quarks. It is also seen from
counting the number of boxes in the Young tableau.

Disregarding the rotation along the 1,2,3 axes (for
example taking only the lowest $J$ state from each band) we
observe from \eq{ErorJE} that at large $N_c$ the spectrum is equidistant
in exoticness,

\beq
{\cal E}_{\rm rot}(E)=\frac{N_c(E+1)}{4I_2},
\la{EE}\eeq
with the spacing $\frac{N_c}{4I_2}=O(1)$. It means that each time we add
a quark-antiquark pair it costs at large $N_c$ the same energy

\beq
{\rm energy\;of\;a\;pair} = \frac{N_c}{4I_2}=O(N_c^0).
 \la{cost}\eeq
Naively one may think that this quantity should be approximately
twice the constituent quark mass. Actually, it can be much less
than that. For example, in the Chiral Quark Soliton Model
\cite{DPPob,Blotz} an inspection of  $I_2$ given there shows that
the pair energy is strictly less than $2M$; in fact $1/I_2$ tends to
zero in the non-relativistic limit of the model. In physical
terms, the energy cost of adding a pair can be small if the pair
is added in the form of a Goldstone boson.

\section{Why collective quantization is valid for exotic multiplets}

\Eq{EE} is interesting as it says that the rotational energy corresponding
to exotic baryons is of the same order as the expected vibrational or radial
excitation energies. Therefore one can suspect that rotational excitations
might mix up with those other ones, and no reliable predictions for exotic
baryons, based on the collective quantization of baryon rotation,
can be made. This doubt has been formulated by Cohen  \cite{Cohen}
who writes: ``The key point is that a collective description is valid only
for motion which is slow compared to vibrational modes which are of order
$N_c^0$ ... The characteristic time scale of quantized collective motion
is given by the typical quantum mechanical result
$\tau \sim (\Delta{\cal E})^{-1}$, where $\Delta {\cal E}$ is the splitting
between two neighboring collective levels." Since the splittings
are of the same order for vibrations and rotations
 he concludes that the rotational description of exotics is not valid
\footnote{In ref. \cite{Cohen} only the splitting of the ``antidecuplet''
has been considered but not the general case.}.

We shall demonstrate that the characteristic time for rotation
is in fact  much larger than Cohen's estimate. Next we calculate the
vibration-rotation mixing and prove that it is small at large $N_c$.

One can find directly the rotation time from angular
velocities. Unusual as it might be, the estimate of the time scale for
rotation from level splitting turns out to be wrong in this particular case:
actually the rotation is much slower.

From \eq{ErotJE} one immediatelly gets the sum of the squares of angular
momenta for a particular ($J,E$) multiplet:
\bea
\la{13}
\sum_{A=1}^3J_A^2&=& J(J+1),\\
\n\\
\la{47}
\sum_{A=4}^7J_A^2 &=& E^2+E(\frac{N_c}{2}+1-J) + \frac{N_c}{2}
\quad\stackrel{N_c\!\to\!\infty}{\longrightarrow}\quad
\frac{N_c}{2}(E+1),
\eea
meaning that
$J_{4\!-\!7}=O(\sqrt{N_c})$ even for zero exoticness. This
should be contrasted to the $SU(2)$ angular momenta:
$J_{1\!-\!3}=O(1)$. For this reason the splittings inside a band
with given $E$ are $O(1/N_c)$ whereas the splitting between the
bands with growing exoticness is $O(1)$.

Let us estimate the angular velocities. As operators,
$\hat{\Omega}_{1\!-\!3}=\hat{J}_{1\!-\!3}/I_1$ and
$\hat{\Omega}_{4\!-\!7}=\hat{J}_{4\!-\!7}/I_2$. Since
positive-definite sums of angular momenta's squares are fixed by
\eq{sums} one obtains
\bea
\la{Omegi13}
\sqrt{<\!\Omega_{1\!-\!3}^2\!>}
&\leq&
\frac{\sqrt{J_1^2+J_2^2+J_3^2}}{I_1}= O\left(\frac{J}{N_c}\right),\\
\n\\
\la{Omegi47} \sqrt{<\!\Omega_{4\!-\!7}^2\!>} &\leq&
\frac{\sqrt{J_4^2+J_5^2+J_6^2+J_7^2}}{I_2}
=O\left(\frac{\sqrt{E+1}}{\sqrt{N_c}}\right).
\eea
We see that angular velocities $\Omega_{4\!-\!7}\gg \Omega_{1\!-\!3}$ at
large $N_c$ even for non-exotic multiplets ($E\!=\!0$). From this point
of view exotic and non-exotic multiplets are on equal footing. It is related
to that all multiplets have high dimensions at large $N_c$. However,
$\Omega_{4\!-\!7}$ {\em are small for any exoticness} $E\!\ll\!N_c$, and the
soliton rotation for such multiplets is slow.

Normally, \eq{EE}  would imply from
the semiclassical Bohr's correspondence principle that the
characteristic time scale for rotation is $\tau\sim (\Delta {\cal
E})^{-1}=O(1)$, as assumed  in ref. \cite{Cohen}. From the explicit
calculation above we see, however, that the characteristic time scale for
rotation is instead $\tau\sim (\Omega_{4\!-\!7})^{-1}=O(\sqrt{N_c})$. This
is a rather rare occasion when the estimate of the time scale from the
semiclassical Bohr's correspondence principle is incorrect. The
reason is that in this particular case the rotation is not
semiclassical because of the quantization condition $J_8 =
-N_c/2\sqrt{3}$. It arises from the Wess--Zumino--Witten term in
\eq{Lrot8} which is a full derivative and, classically, should be
thrown out.

The characteristic time scale for rotation leading to exotic
multiplets is thus large, $\tau=O(\sqrt{N_c})$. In particular,
rotation is much slower compared to the vibrational or radial
modes whose excitation energies are $O(1)$ and hence their time
scale is $O(1)$. Therefore, it is fully legitimate to treat the
rotation leading to exotic multiplets as collective excitations of
a chiral soliton, at least as long as $E\!\ll\! N_c$.

To make this statement quantitative, we derive the shift in the exotic
baryons' rotation energy  as due to mixing with non-rotational degrees of
freedom. We show that this shift  is $O(1/N_c)$ and tends to zero at large
$N_c$. It is a direct proof that exotic multiplets can be treated as
collective excitations. We first present a general derivation of the mixing
and then give a more transparent quantum mechanical example with only one
non-rotational degree of freedom but preserving all properties of the case
at hand. To purify the argument we imply here an idealized case of
$N_c\!\to\!\infty$ and neglect non-zero $m_s$.\\

\newpage
\noi
\underline{General chiral soliton derivation}\\

At large $N_c$ baryons are chiral solitons \cite{Witten} corresponding to
a local minimum of some effective chiral action $S_{\rm eff}[\pi(x)]$. Its
specific form is irrelevant to the argument: the important thing is that
it is proportional to $N_c$. Its local minimum $\pi_{\rm class}(x)$
gives the soliton profile, and the moments of inertia $I_{1,2}$ are computed
at this minimum. Hence ${\cal M}_0$, $I_{1,2}$ are all proportional to
$N_c$.

Vibrational modes of a baryon are encoded in quantum fluctuations about
the classical minimum. One expands the static energy part of the effective
chiral action about the minimum:

\beq
{\cal E}_{\rm eff}[\pi_{\rm class}+\delta\pi] =
{\cal M}_0+\frac{1}{2}\delta\pi\,W[\pi_{\rm
class}]\,\delta\pi+\ldots
\la{S2}\eeq
where $W$ is some operator in a given external field $\pi_{\rm class}$ and
is usually referred to as the quadratic form. It is of the order of $N_c$
(since the full $S_{\rm eff}$ is), hence quantum fluctuations scale as
$\delta\pi(x) =O(1/\sqrt{N_c})$. The spectrum and eigenfunctions of $W$
are $N_c$-independent. $W$ has zero modes related to symmetry, in this case
translations and rotations. The quantization of rotations (which are large
fluctuations as they occur in flat zero-mode directions) leads to the
rotational spectrum discussed in the previous section. The vibrational modes
are orthogonal to those zero modes. One can expand a general fluctuation in
eigenfunctions of the quadratic form:
\bea
\la{eig}
W\psi_n(x)&=&\kappa_n\psi_n(x),\qquad \kappa_n>0,\\
\la{Four}
\delta\pi(x)&=&\sum_n c_n\psi_n(x).
\eea
Assuming eigenfunctions are (ortho) normalized to unity, the Fourier
coefficients are $c_n=O(1/\sqrt{N_c})$ and the eigenvalues are
$\kappa_n=O(N_c)$ since $W$ is proportional to $N_c$.  $c_n$'s can be
considered as normal coordinates for vibrations. In the harmonic
approximation their hamiltonian is

\beq
{\cal H}_{\rm vibr} =
\sum_n\left(-\frac{1}{2\mu_n}\,\frac{\partial^2}{\partial c_n^2} +
\frac{\kappa_n}{2}c_n^2\right)
\la{Hharm}\eeq
with $\kappa_n,\mu_n=O(N_c)$ leading to vibration energies
$\epsilon_{n,k_n}=\sqrt{\kappa_n/\mu_n}(k_n+\half)=O(1)$, as it should be.
The ground-state ($k_n\!=\!0$) wave function is $\psi(c_n)=
\exp(-\sqrt{\kappa_n\mu_n}c_n^2/2)$.

We now consider the influence of vibrations on the rotational spectrum
${\cal E}_{\rm rot}=\frac{N_c(E+1)}{4I_2[\pi]}$. In the leading
(classical) order one substitutes the classical soliton field and gets the
moment of inertia $I_{20}=I_2[\pi_{\rm class}]=O(N_c)$. Taking into account
quantum fluctuations one expands

\beq
I_2[\pi_{\rm class}+\delta\pi]=I_{20}+\delta I_2
+\delta^2 I_2+\ldots
=I_{20}+\sum\alpha_nc_n
+\sum_{m,n}\beta_{mn}c_mc_n+\ldots,\qquad
\alpha_n,\beta_{mn}=O(N_c).
\la{I2expan}\eeq
Consequently, the hamiltonian for rotation-vibration mixing is

\beq
{\cal H}_{\rm rot-vibr}=\frac{N_c(E+1)}{4I_2[\pi_{\rm class}+\delta\pi]}
=\frac{N_c(E+1)}{4I_{20}}\left[1-\frac{\sum_n\alpha_nc_n}{I_{20}}
+\frac{\sum_{m,n}(\alpha_m\alpha_n-\beta_{mn}I_{20})c_mc_n}{I_{20}^2}
+\ldots\right].
\la{Hrotvibr}\eeq
One can now evaluate the corresponding rotation-vibration mixing energy
by perturbation theory in quantum fluctuations $c_n$. The term linear in
$c_n$ is zero in the first order but in the second order perturbation theory
it is non zero and one gets

\beq
{\cal E}^{(1)}_{\rm rot-vibr}\;\simeq\;
{\cal E}_{\rm rot}^2 \frac{<\!1|\sum \alpha_nc_n|0\!>^2}{\Delta{\cal
E}_{\rm vibr}\,I_{20}^2}\;\simeq\;
{\cal E}_{\rm
rot}^2\sum_n\frac{1}{\kappa_n} \left(\frac{\alpha_n}{I_{20}}\right)^2
\;\simeq\; {\cal E}_{\rm rot}\,\frac{{\cal E}_{\rm rot}}{\Delta{\cal
E}_{\rm vibr}}\,\left<\left(\frac{\delta
I_2}{I_2}\right)^2\right>\;\propto\; O\left(\frac{(E+1)^2}{N_c}\right)
\la{Emix1}\eeq
which is a small $1/N_c$ correction to the main rotational energy,
\eq{EE}. Only when exoticness is comparable to $N_c$ the correction becomes
of the same order as the main term.

Another contribution to the mixing arises from the last term in
\eq{Hrotvibr}. Here it is sufficient to use the first order perturbation
theory, and one obtains

\beq
{\cal E}^{(2)}_{\rm rot-vibr}={\cal E}_{\rm rot}
\sum_n\frac{1}{\sqrt{\kappa_n\mu_n}
}\frac{\alpha_n^2-\beta_{nn}I_{20}}{I_{20}^2}
\;\;\simeq \;\; {\cal E}_{\rm rot}\,\left<\left(\frac{\delta
I_2}{I_2}\right)^2-\frac{\delta^2I_2}{I_2}\right> \;\;\propto\;\;
O\left(\frac{E+1}{N_c}\right).
\la{Emix2}\eeq
This term is a small correction to ${\cal E}_{\rm rot}$ \ur{EE} even at
large exoticness. Thus we have proved that, despite the rotational
energy for exotic baryons being $O(1)$, its mixing with vibrational degrees
of freedom leads to a small $O(1/N_c)$ correction.

The change in the baryon form owing to rotation cannot be
neglected only when the rotational energy reaches
${\cal E}_{\rm rot} = O(N_c)$ comparable to the static baryon mass
${\cal M}_0$. At this point everything goes wrong \cite{BR,DPRegge}: the
widths become comparable to the masses owing to strong pion radiation and
the centrifugal forces deform the baryon such that the rotational
energy has to be computed anew. \\

\noi
\underline{Charged particle in the field of a monopole}\\

To illustrate the general derivation above, we consider  a
quantum-mechanical example preserving the equidistant rotational spectrum
with the spacing of the same order as in the vibrational one. In fact we
take the example of Guadagnini \cite{Guad} which he used as analogy to
derive the quantization of the $SU(3)$ skyrmion. This example is an ideal
copy of the case at hand, except that there is only one non-rotational
degree of freedom instead of an infinite number as in the real case.
At the same time it is so simple that all calculations can be done
explicitly and to any given accuracy.

Consider a charged particle on a sphere of radius $R$ surrounding
a magnetic monopole whose magnetic field is ${\bf B}={\bf r}/r^3$.
The lagrangian can be written in terms of the angular velocities
of the particle $\Omega_{1,2,3}$ with a ``Wess--Zumino--Witten"
term linear in $\Omega_3$ \cite{Guad,DP7}:

\beq
{\cal L}_{\rm rot}=\frac{I}{2}(\Omega_1^2+\Omega_2^2)+eg\, \Omega_3,
\la{Lmon}\eeq
where $eg$ is the product  of the electric  and magnetic charges; it must be
an integer. The canonical quantization leads to the hamiltonian written in
terms of orbital momentum operators

\beq
{\cal H}_{\rm rot}=\frac{1}{2I}(L_1^2+L_2^2)
\la{Hmon}\eeq
supplemented with the quantization condition $L_3=eg$, similar
to what one requires in \eq{Hrot}. The hamiltonian has eigenvalues
${\cal E}_{\rm rot}=[L(L+1)-(eg)^2]/2I$. At large $N=eg$ (being a direct
analog of large $N_c$) we introduce an analog of exoticness
$E\equiv L-N$ and rewrite the rotational energy as

\beq
{\cal E}_{\rm rot} = \frac{E^2+E(2N+1)+N}{2I}.
\la{Emon}\eeq
This is fully similar to \eq{ErotJE}: at large $N$ but
finite $E$ the rotational spectrum is equidistant with the spacing
$\Delta {\cal E}=N/I$, and we imply that the moment of inertia is
$I=\mu R^2=O(N)$, with $R=O(1)$ and particle mass $\mu=O(N)$.

To study the interplay between rotations and vibrations, we now allow the
particle to deviate from the sphere of radius $L$ putting it in a potential
well $\kappa(|{\bf r}|-R)^2/2$. We arrive at the hamiltonian \cite{Guad}

\beq
{\cal H} =
-\frac{1}{2\mu}\left(\frac{d^2}{dr^2}+\frac{2}{r}\frac{d}{dr}\right)+
\frac{(2E+1)N}{2\mu r^2}+\frac{\kappa(r-R)^2}{2}
\la{HmonV}\eeq
where we take $\kappa=O(N)$ to have vibrational excitations stable in $N$.

If one neglects the fluctuations $x\equiv r-R$ as compared to the average
distance $R$, the rotation and vibration variables are completely separated
and one obtains
\bea
\n
{\cal E}_{\rm rot}(E)&=&\frac{(2E+1)N}{2I}=O(1),\qquad {\rm
moment\;of\;inertia}\;I=\mu R^2,\\
\n\\
\n
{\cal E}_{\rm
vibr}(k)&=&\sqrt{\frac{\kappa}{\mu}}\left(k+\frac{1}{2}\right)
=O(1),\qquad
\psi(x) = \exp\left(-\sqrt{\kappa\mu}\,\frac{x^2}{2}\right),\\
\n\\
\n
<\!x^2\!> &\simeq & \frac{1}{\sqrt{\kappa\mu}} =
O\left(\frac{1}{N}\right).
\eea
The interaction between rotations and vibrations comes through the
centrifugal term in \eq{HmonV}. We expand
$1/(R+x)^2=1/R^2(1-2x/R+3x^2/R^2+\ldots)$ and estimate it
using the harmonic oscillator wave functions in $x$. The leading term is
zero by parity, therefore we have to apply the second-order perturbation
theory for the $x$ term and the first-order perturbation theory for the
$x^2$ term. The former gives

\beq
{\cal E}_{\rm rot-vibr}^{(1)}\;\;\simeq\;\;  {\cal E}_{\rm rot}\,\frac{{\cal
E}_{\rm rot}}{\Delta{\cal E}_{\rm vibr}}\frac{<\!1|x|0\!>^2}{R^2}
\;\;\simeq\;\;
{\cal E}_{\rm rot}^2\,\frac{1}{\kappa R^2}\;\;\propto\;\;
\frac{(2E+1)^2}{N}.
\la{Erotvibr11}\eeq
The latter gives

\beq
{\cal E}_{\rm rot-vibr}^{(2)}\;\;\simeq\;\;  {\cal E}_{\rm
rot}\,\frac{<\!x^2\!>}{R^2}
\;\;\simeq\;\;  {\cal E}_{\rm rot}\,\frac{1}{R^2\sqrt{\kappa\mu}}\;\;
\propto\;\; \frac{2E+1}{N}.
\la{Erotvibr21}\eeq
It is always a small correction. As to \eq{Erotvibr11},  it becomes a
sizable correction only when the `exoticness' compares to $N$,
like in the above general analysis.

In this example the rotational splittings are themselves $O(1)$, like the
splittings between baryons with different exoticness. Nevertheless, the
``baryon masses'' are computed accurately from the rotational spectrum.
Furthermore, one sees that the rotational level
spacing is actually irrelevant to the shift of the rotational energy owing
to vibrations. The only thing which counts is the {\em change of
the moment of inertia}. This change is $\delta I/I=O(1/N_c)$ in
the general case and in the example considered. The change $\delta I/I$
becomes of the order of unity in two cases: {\it i}) when the vibrational
excitation is of the order of $N_c$ so that $\delta x/R = O(1)$,
{\it ii}) when rotational energy is of the order of $N_c$ so that
${\cal E}_{\rm rot}/{\cal E}_{\rm vibr}=O(N_c)$. In such cases
the deformation of the soliton  is not small, and one cannot
consider it as a rigid body. \\

Finally, let us comment on the recent suggestion \cite{Cohen} that
$\Theta^+$ could be extracted from a linear response theory describing
meson-soliton scattering \cite{HEHW,MK}. Essentially the same idea is
proposed in ref. \cite{CohLeb} to find $\Theta^+$ partner states, using
the chiral-soliton formalism of ref. \cite{MM}.

The spectrum of vibrational modes is defined by the quadratic form
$W$. In a specific (Skyrme) model this spectrum has been studied
in refs. \cite{HEHW,MK}. The spectrum is naturally
$N_c$-independent, as are the ensuing meson-baryon scattering
amplitudes. However it does not mean that {\em all}
$N_c$-independent excitations will be seen as poles in the
scattering amplitudes generated by the quadratic form. It is well
known that poles related to rotational excitations  are missed in
the quadratic form: they correspond to fluctuations in flat
zero-mode directions, and cannot be considered as small. All
rotational states, independently of their energy, arise as poles
in meson-baryon scattering amplitudes from Born graphs  which are
missing in the quadratic form but have been recovered in ref.
\cite{DPPBorn} as a purely classical effect. In particular,
assuming $\Theta^+$ is a $O(1)$ rotational excitation, it will not
occur in the small-oscillation spectrum, contrary to the statement
of refs. \cite{Cohen,CohLeb}.

\section{Summary}

We have constructed the generalization of $\oct$, $\dec$, $\adec$,
$\pletthree$, $\pletone$... multiplets to the case of arbitrary
$N_c$. These multiplets are classified by ``exoticness" -- the
number of extra quark-antiquark pairs needed to compose the
multiplet. The splittings between masses of the multiplet with the
same exoticness are $O(1/N_c)$ {\it i.e.} parametrically small at
large $N_c$. The spectrum of multiplets with growing exoticness
is equidistant with an $O(1)$ spacing. On the one hand this spacing
can be interpreted as an energy for adding a quark-antiquark pair
but on the other hand it is a rotational excitation. Despite that it becomes
comparable to vibrational or radial excitations of baryons, both non-exotic
and exotic bands are, at large $N_c$, reliably described as collective
excitations of the ground-state baryons: the corrections die out as
$1/N_c$. This conclusion is opposite to the recent claim of ref.
\cite{Cohen}. That claim has been based on an assumption that the rotation
corresponding to exotic baryons is fast but in fact it is slow, as we have
explicitly shown. The collective quantization description
fails only when the exoticness becomes comparable to $N_c$.

The newly discovered $\Theta^+$ baryon belongs to
the exoticness=1 multiplet $\adec$. The larger $N_c$ the
more accurate would be its description as a rotational state
of a chiral soliton.

\vskip .5true cm

We are grateful to Pavel Pobylitsa and Maxim Polyakov for
many fruitful discussions. The figures in this paper have been
prepared using the graphical editor ``Feynman'' designed by
Pobylitsa.  We thank Thomas Cohen and Richard Lebed for a
correspondence. V.P. acknowledges hospitality at Nordita and
a partial support by the grant RFBR-0015-9606.

\end{document}